\documentclass[aps,prd,a4paper,reprint,nofootinbib,superscriptaddress,floatfix,preprintnumbers,amsmath,amssymb]{revtex4-2}
\usepackage[T1]{fontenc}
\usepackage[utf8]{inputenc}
\usepackage[UKenglish]{babel}
\usepackage{mathtools,hyperref,tikz,physics,subcaption,dsfont}
\usepackage{booktabs}
\usepackage{array}
\usepackage[justification=raggedright]{caption}
\newcommand{\oo}{\mathrm{O}}
\newcommand{\oob}{\bar{\mathrm{O}}}
\newcommand{\zzero}{\mathbf{0}}

\newcommand{\appendixsection}{
	\setcounter{equation}{0}
	\setcounter{section}{0}
	\setcounter{figure}{0}
	\renewcommand{\theequation}{S\arabic{equation}}
	\renewcommand{\thefigure}{S\arabic{figure}}
	\onecolumngrid
	\vspace*{.7cm}
	\hrule
	\vspace*{.04cm}
	\hrule
	\begin{center}
		\vspace*{.4cm}
		{\bf \large Supplemental Material}
		\vspace*{.5cm}
	\end{center}
	\twocolumngrid
}

\begin{document}
\begin{flushright}
    \texttt{DESY-24-202}
\end{flushright}
\title{Nucleon sigma terms with a variational analysis from Lattice QCD}
\author{Lorenzo Barca}
\email{lorenzo.barca@desy.de}
\affiliation{John von Neumann-Institut für Computing (NIC), Deutsches Elektronen-Synchrotron (DESY), Platanenallee 6, 15738 Zeuthen, Germany}

\author{Gunnar Bali}
\email{gunnar.bali@ur.de}
\affiliation{Fakultät für Physik, Universität Regensburg, Universitätsstraße 31, 93053 Regensburg, Germany}

\author{Sara Collins}
\email{sara.collins@ur.de}
\affiliation{Fakultät für Physik, Universität Regensburg, Universitätsstraße 31, 93053 Regensburg, Germany}

\date{\today}

\begin{abstract}
We determine the nucleon-sigma terms from lattice QCD. We find that
the dominant excited state contamination in the nucleon three-point
function with a scalar current is due to the transition between the
nucleon and a S-wave scattering state of a nucleon and a scalar
(sigma) meson.  In this proof-of-concept study, we analyse a single
$N_f=3$ ensemble with the unphysically large pion mass
$M_\pi=429$ MeV. Excited state contamination is substantially reduced
compared to the standard method when employing nucleon-sigma type
interpolating operators within a generalised eigenvector analysis.
\end{abstract}
\maketitle

\section{Introduction}
Almost all visible matter of the universe is composed of nucleons,
i.e.\ protons and neutrons. Most of the nucleons' mass can be attributed
to the spontaneous breaking of chiral symmetry. Only a small part is
due to the Higgs mechanism, i.e.\ the masses of the valence and sea
quarks.  These contributions can be quantified in a scheme- and
gauge-invariant way in terms of scalar matrix elements, the
quark-nucleon $\sigma$-terms,
\begin{equation}
  \sigma_{qN}=m_q\langle N|\bar{q}{q}|N\rangle
  =  m_q\frac{\partial m_{\mathrm{N}}}{\partial m_q},
  \label{eq:sigma}
\end{equation}
where $q\in\{u,d,s,\ldots\}$. These determine the coupling strength of
the Standard Model Higgs boson (or other scalar particles) at zero
recoil to the nucleon and are therefore important for theory
predictions relevant for the direct detection of weakly interacting
massive dark matter particles, see, e.g., the
review~\cite{Alarcon:2021dlz}.

In the isospin symmetric limit, i.e.\ for equal up and down quark
masses $m_u=m_d$ and electric charges, the pion-nucleon $\sigma$-terms
$\sigma_{\pi{}N}=\sigma_{uN}+\sigma_{dN}$ are the same for the proton
and the neutron: $\sigma_{\pi p}=\sigma_{\pi n}$. $\sigma_{\pi{}N}$
can be extracted indirectly via dispersive analyses of pion-nucleon
scattering data, employing low energy theorems and chiral perturbation
theory (ChPT). Initially, a value
$\sigma_{\pi{}N}=45(8)$~MeV~\cite{Gasser:1990ce} was obtained.  More
recent phenomenological determinations, when corrected for isospin
breaking effects~\cite{Hoferichter:2023ptl}, scatter around
56~MeV~\cite{Alarcon:2011zs,Chen:2012nx,Hoferichter:2015dsa,RuizdeElvira:2017stg,Hoferichter:2023ptl}. The
$\sigma$-terms can also be determined from first principles lattice
QCD, either
directly~\cite{Yang:2015uis,Bali:2016lvx,Yamanaka:2018uud,Gupta:2021ahb,Agadjanov:2023efe, Alexandrou:2024ozj}
or
indirectly~\cite{Alexandrou:2014sha,Durr:2015dna,Borsanyi:2020bpd,RQCD:2022xux,
  Hu:2024mas} from the quark- (or the pseudoscalar meson mass-)
dependence of the nucleon mass, using the second equality of
Eq.~\eqref{eq:sigma}.  The most recent Flavour Lattice Averaging Group
(FLAG) average of results obtained via the direct and the indirect
methods, reads for the $N_f=2+1$ theory
$\sigma_{\pi{}N}=42.2(2.4)$~MeV~\cite{FlavourLatticeAveragingGroupFLAG:2024oxs}.
No tension is seen between the individual determinations. In contrast,
the $N_f=2+1+1$ average reads
$\sigma_{\pi{}N}=60.9(6.5)$~MeV~\cite{FlavourLatticeAveragingGroupFLAG:2024oxs}.
This average, which is closer to the recent phenomenological
estimates, is dominated by the PNDME result~\cite{Gupta:2021ahb},
obtained via the direct method.

Nucleon matrix elements, including the scalar matrix elements, are
extracted from combinations of Euclidean two- and three-point
functions. Within these, interpolating operators (interpolators) with
the quantum numbers of the nucleon at a given three-momentum are used
to create and to destroy hadronic states that propagate in Euclidean
time.  In the three-point function, a local quark bilinear current is
inserted at an intermediate time. The interpolators not only create
the ground state of interest but also excited states, including a
tower of multi-particle states composed of a baryon and mesons.  The
ground state properties of interest can in principle be obtained in
the limit of large Euclidean time separations between the ``source'',
the current (in the case of three-point functions) and the ``sink''.
Due to the exponential noise over signal problem of baryonic Green
functions~\cite{Hamber:1983vu,Lepage:1989hd}, excited state
contributions are often still significant for the separations that are
achievable within reasonable precision.  These are then accounted for
by carrying out multi-exponential fits. Within three-point functions,
the transitions between the ground and some excited states can be
particularly enhanced since the current can directly create a meson
which, in combination with a baryon, can give the quantum numbers of
the final state nucleon.  The transition from $N$ states to $N\pi$ and
$N\pi\pi$ states, mediated by a current, has been investigated using
ChPT~\cite{Tiburzi:2009zp,Bar:2016uoj,Bar:2018xyi,Bar:2021crj,Gupta:2021ahb}.
Results of such calculations have then inspired the fit ansätze used
in lattice analyses, e.g., of axial and pseudoscalar
form factors~\cite{RQCD:2019jai,Jang:2023zts} and
$\sigma$-terms~\cite{Gupta:2021ahb,Agadjanov:2023efe}.

Motivated by ChPT, PNDME~\cite{Gupta:2021ahb} constrain the gap
between the ground and the first excited state energies to agree
within 20\% with the difference between the nucleon mass and the
lowest possible non-interacting $N\pi$ (P-wave) or $N\pi\pi$ (S-wave)
energy level. In their multi-state analysis this has a large impact, in
particular, at small pion masses and for the disconnected quark line
Wick contraction, resulting in the final value
$\sigma_{\pi{}N}=59.6(7.4)$~MeV. This is larger than what was found
in other recent direct
determinations~\cite{Yang:2015uis,Bali:2016lvx,Yamanaka:2018uud,Agadjanov:2023efe, Alexandrou:2024ozj} 
or predictions using the indirect (Feynman-Hellmann)
method~\cite{Durr:2015dna,Borsanyi:2020bpd,RQCD:2022xux}, with
different systematics related to excited states.

A more systematic approach would be to explicitly resolve and subtract
the dominant excited state contributions using a variational approach
with a basis of interpolators, including baryon-meson type
operators. The efficacy of this approach was first demonstrated
in~\cite{Barca:2022uhi} at unphysically large pion masses, where the
contributions from the lowest $N\pi$ excited states were removed from
axial and pseudoscalar three-point functions with and without momentum
transfer.\footnote{No improvement was seen in the forward limit, in
the rest frame.}  Subsequently, in~\cite{Alexandrou:2024tin} this was
confirmed at the physical pion mass and extended to vector, tensor and
scalar matrix elements. In the latter case no improvement was found
when adding the lowest P-wave nucleon-pion type operator to the
interpolator basis. This may not be surprising since one would expect
flavour-diagonal scalar currents to dominantly couple to isoscalar
scalar meson resonances. At the physical point this would be the
$f_0(500)$ (decaying into $\pi\pi$) and the $f_0(980)$.  In terms of
the flavour content, these are mixtures of the scalar $I=I_3=S=0$
singlet and octet SU(3) eigenstates $\sigma_0$ and $\sigma_8$,
respectively.

Restricting ourselves to the baryon-(single)meson sector, a number of 
interpolators with the quantum numbers of a nucleon at rest can be
constructed. These include P-wave combinations of type $N\pi$,
$N\eta_0$, $N\eta_8$, $\Sigma K$ etc.\ as well as S-wave combinations
like $Na_0$, $N\sigma_0$, $N\sigma_8$, $\Sigma K_0^*$ and so on. To
resolve the dominant multi-particle excited state it may be sufficient
just to employ interpolators containing flavour-diagonal scalar mesons
since only these can be created directly by the $\bar{u}u+\bar{d}d$
and $\bar{s}s$ currents of interest. These are the $\sigma_0$ and
$\sigma_8$ mesons with flavour content
$(\bar{u}u+\bar{d}d+\bar{s}s)/\sqrt{3}$ and
$(\bar{u}u+\bar{d}d-2\bar{s}s)/\sqrt{6}$, respectively.  In addition
we consider a $N\sigma$ interpolator with
$\sigma\sim(\bar{u}u+\bar{d}d)/\sqrt{2}$.

In this proof-of-concept study, we investigate whether excited state
contamination for scalar matrix elements can indeed be reduced in a
variational approach, including either $N\sigma$ or $N\sigma_0$ and
$N\sigma_8$ type interpolators in the basis.  This work is carried out
on a single Coordinated Lattice Simulations~\cite{Bruno:2014jqa}
ensemble employing $N_f=3$ non-perturbatively improved Wilson fermions
at a larger than physical pion mass\footnote{The trace of the mass
matrix $m_u+m_d+m_s$ is close to that at the physical point.}
$M_\pi\approx 429$~MeV on the SU(3) flavour symmetric line
($m_q=m_u=m_d=m_s$, i.e., $M_K=M_\pi$) and a lattice spacing $a\approx
0.098$~fm. For this quark mass combination the $\sigma_0$ meson is
stable with the mass $M_{\sigma_0}=554(49)$~MeV (see the Supplemental
Material), similar to the real part of the pole of the $f_0(500)$
observed in nature. We remark that the variational approach with a basis 
of $N$ and $N\sigma$ type interpolators~(along with, e.g., $N\pi$ 
interpolators) has been investigated previously in the context of 
scattering studies~\cite{Lang:2016hnn,Kiratidis:2016hda}.

This Letter is organised as follows. In Sec.~\ref{sec:method} we
introduce the general procedure for the direct determination of the
renormalized $\sigma$-terms. We then detail our variational
determination of the energy eigenvalues and eigenvectors
in Sec.~\ref{sec:variat} and compare results from fits to the
improved and standard Green functions in Sec.~\ref{sec:fit},
before we conclude in Sec.~\ref{sec:conclude}.

\begin{figure*}[t]
  \centerline{
    \includegraphics[width=0.98\textwidth]{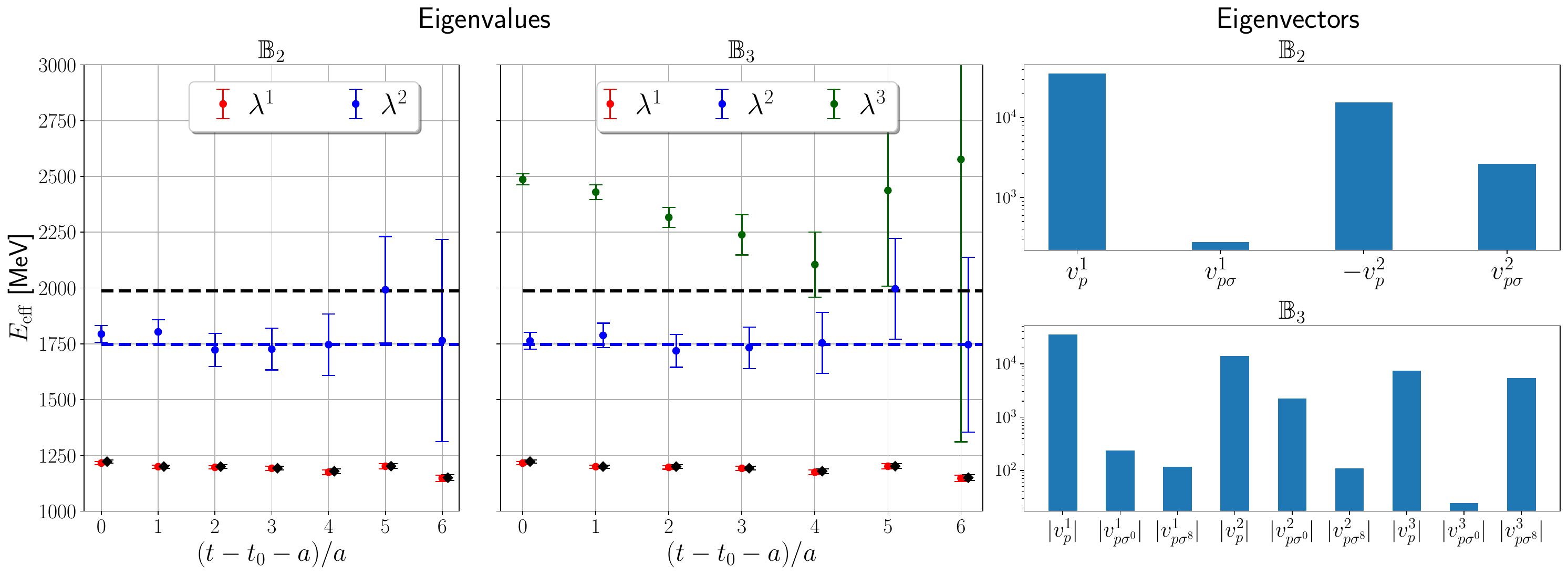}}
  \caption{\label{gevp_results}Effective energies obtained from the
    GEVP with interpolator bases $\mathbb{B}_2$~(left) and
    $\mathbb{B}_3$~(middle) for $t_0=3a$, compared to the effective
    mass of the standard nucleon two-point function (black diamonds)
    constructed from the $\oo_p$ operators.  The dashed lines
    correspond to the energies of the non-interacting S-wave
    $N\sigma$~(blue) and lowest P-wave $N\pi$~(black) levels.
    Also shown are the corresponding generalised eigenvector 
    components for $t-t_0=3a$ for $\mathbb{B}_{2}$~(right top) 
    and $\mathbb{B}_{3}$~(right bottom).}
\end{figure*}

\section{Direct determination of the $\sigma$-terms}\label{sec:method}
The scalar matrix elements can be extracted from two- and three-point
functions
\begin{align}
\label{NSN_3pt}
C_{\rm 2pt}(t) &= \langle \oo(t) \oob(0)  \rangle,\\
C^{\mathcal{S}^q}_{\rm 3pt}(t, \tau) &= \langle  \oo(t) \mathcal{S}^q(\tau) \oob(0)  \rangle,
\end{align}
in the rest frame. The interpolating operator $\oob$~($\oo$)
creates~(annihilates) states with the quantum numbers of the nucleon
$I(J^P)=\tfrac{1}{2}(\tfrac{1}{2}^+)$ at the source on time slice 0~(sink
on time slice $t$). For the three-point function, the scalar current
$\mathcal{S}^q=\bar{q}q$ is inserted at an intermediate time $\tau$,
where we consider the flavour combinations
$\mathcal{S}^{u+d}=\bar{u}u+\bar{d}d$ and $\mathcal{S}^s=\bar{s}s$.
The spectral decomposition of the correlation functions reads
\begin{align}
C_{\rm 2pt}(t) &= \sum_n |Z_{n}|^2 e^{-E_n t},\label{eq:specdecomp2}\\
C^{\mathcal{S}^q}_{\rm 3pt}(t, \tau) &= 
\sum_{n',n} Z_{n'}\bar{Z}_n \bra{n'} \mathcal{S}^q \ket{n} e^{-E_{n'}(t-\tau)} e^{-E_n\tau}.\label{eq:specdecomp3}
\end{align}
$E_n$ is the energy of state $|n\rangle$, created by the interpolator
$\oob$ from the vacuum state $|\Omega\rangle$, and $\bar{Z}_{n}$ is the
associated overlap factor $\bar{Z}_{n}\propto
\bra{n}\oob\ket{\Omega}$.  In the limit of large time separations
$t\gg\tau\gg 0$, the ground state contribution~($n=n'=N$) dominates.
By performing fits to combinations of the two- and three-point
functions~(see Sec.~\ref{sec:fit}) the nucleon scalar matrix elements
in the lattice scheme can be extracted:
\begin{align}
  \langle N| \mathcal{S}^{u+d}|N\rangle &= \bar{u}_N g_S^{{\rm lat},u+d} u_N, \\
  \langle N| \mathcal{S}^{s}|N\rangle &= \bar{u}_N g_S^{{\rm lat},s} u_N. 
\end{align}
The Lorentz decomposition of these matrix elements is expressed in
terms of the scalar charges $g_S^{{\rm lat},q}$, $q\in\{u,d,s\}$, and
$u_N$, the spinor of a nucleon at rest. For Wilson fermions, on the
SU(3) flavour symmetric line the renormalised $\sigma$-terms are given
by
\begin{align}
  \sigma_{\pi N} &= m_q\left[r_mg_S^{{\rm lat},u+d}+ \frac{2}{3}(1-r_m) g_S^{{\rm lat},u+d+s}\right],\label{eq:sigm1}\\
  \sigma_{s N} &= m_q\left[r_mg_S^{{\rm lat},s}+\frac{1}{3}(1-r_m) g_S^{{\rm lat},u+d+s}\right],\label{eq:sigm2}
  \end{align}
where $m_q$ is the vector Ward identity quark mass and $r_m$ is the
ratio of the singlet to non-singlet mass renormalisation factors.  In
our case, $r_m = 3.409(57)$~\cite{Heitger:2021bmg} and
$m_q=6.62(25)$~MeV~\cite{RQCD:2022xux}.

\section{Variational analysis}\label{sec:variat}

To construct an interpolator with large overlap to the nucleon ground
state, we perform a variational analysis~\cite{Bulava:2011yz}. We
utilise two bases of interpolators with proton quantum numbers
($I_z=+1/2$).  Basis $\mathbb{B}_{2} = \{\oo_p,\oo_{p\sigma}\}$
comprises the usual proton interpolator $\oo_p = u^T(dC\gamma_5 u)$ and
a S-wave proton-sigma meson interpolator $\oo_{p\sigma}$, combining
$\oo_p$ with the SU(2) sigma meson operator~($I=0$, $J^{PC}=0^{++}$),
$\oo_\sigma = \left( \bar{u}u + \bar{d}d\right) / \sqrt{2}$.  For
basis $\mathbb{B}_{3} = \{\oo_p, \oo_{p\sigma_0}, \oo_{p\sigma_8}\}$,
we implement the SU(3) singlet and octet sigma meson operators,
$\oo_{\sigma_0} = \left(\bar{u}u + \bar{d}d + \bar{s}s\right) /
\sqrt{3}$ and $\oo_{\sigma_8} = \left(\bar{u}u + \bar{d}d -2
\bar{s}s\right) / \sqrt{6}$, respectively.  To improve the overlap
with lower-lying levels, we construct spatially extended interpolators
by Wuppertal smearing~\cite{Gusken:1989ad,Gusken:1989qx} the quark
fields, incorporating APE smeared links~\cite{Falcioni:1984ei}. \
For the nucleon and $\sigma$ operators within the single- and
two-hadron interpolators we employ the smearing radii
$\langle r^2\rangle^{1/2}\sim 1.0$~fm and $\langle
r^2\rangle^{1/2}\sim 0.2$~fm, respectively.

\begin{figure*}[htb]
	{\centering
	\begin{subfigure}[b]{0.98\textwidth}
		\centering
		\includegraphics[width=\textwidth]{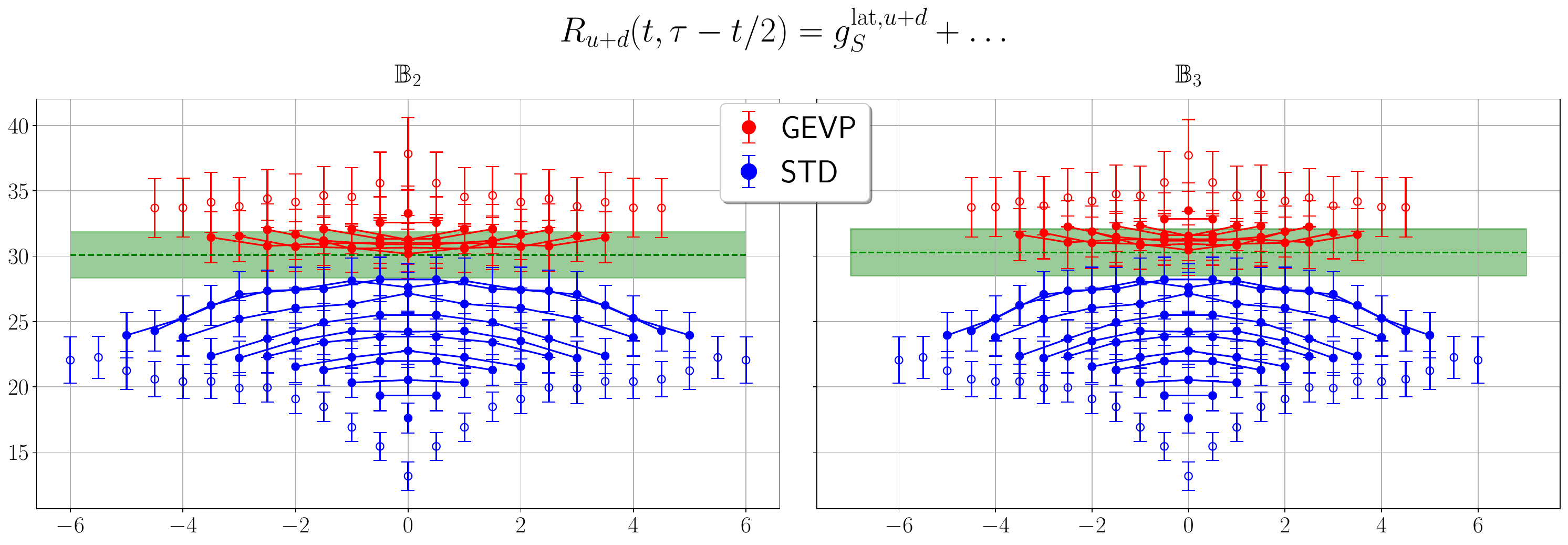}
	\end{subfigure}
	\begin{subfigure}[b]{0.98\textwidth}
		\centering
		\includegraphics[width=\textwidth]{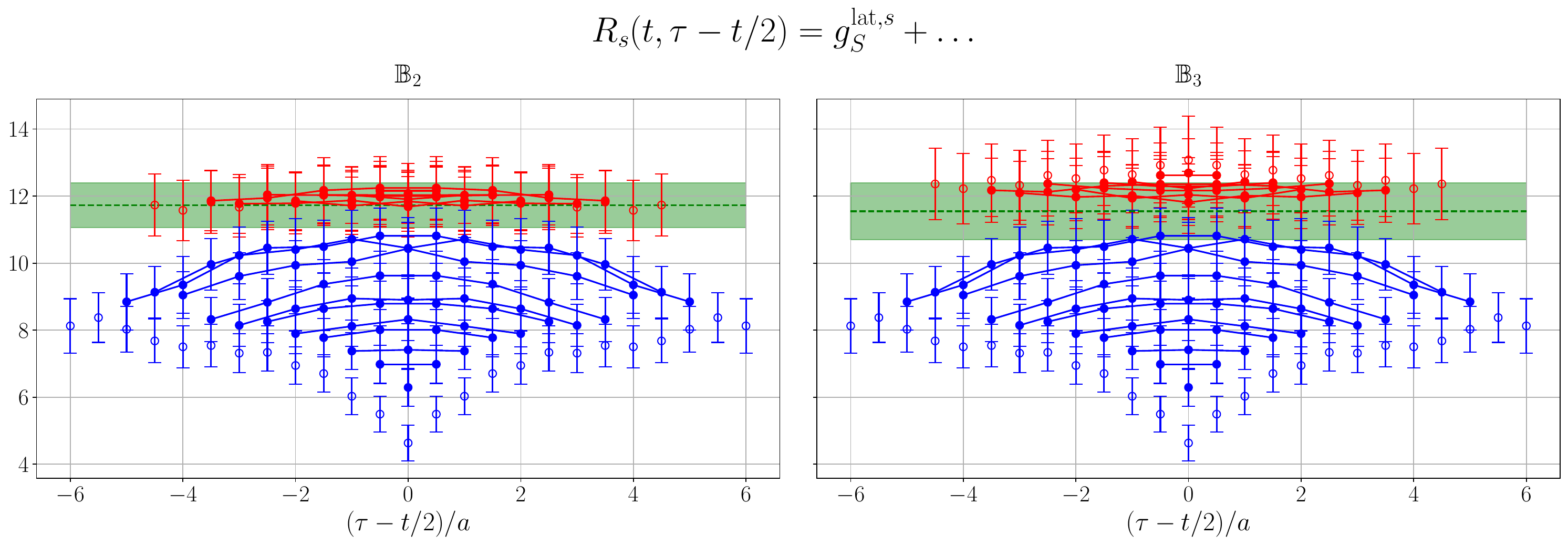}
	\end{subfigure}}
	\caption{\label{fig:gevp_std_ratios}Results for the
          GEVP-improved ratios $R_{u+d}$~(top) and $R_s$~(bottom) for
          $\mathbb{B}_2$ (left) and $\mathbb{B}_3$ (right). The green
          dashed lines and bands show the central values and
          errors, respectively, of $g^{{\rm lat},u+d}_S$~(top) and
          $g^{{\rm lat},s}_S$~(bottom) extracted using the summation
          method, see Sec.~\ref{sec:fit}. The standard ratios obtained
          using the interpolator $\oo_p$ are shown for comparison. The
          data points that are utilised in the fits (Sec.~\ref{sec:fit})
          are indicated by solid circles.}
\end{figure*}

Using these bases, we build the following matrices of two-point functions:
\begin{align}\label{gevp_c2pt}
&\mathbb{C}_{\rm 2pt}(t)_{ij}
=
\langle \oo_i(t) \oob_j(0) \rangle 
\end{align}
with $\oo_i, \oo_j \in \mathbb{B}_2$ or $\mathbb{B}_3$.  The
corresponding Wick contractions are evaluated using the sequential
source method~\cite{Maiani:1987by} for quark-line connected
topologies.  For the disconnected diagrams also stochastic estimation
is used, including the
one-end-trick~\cite{Sommer:1994gg,Foster:1998wu}, see the Supplemental
Material. We solve the generalised eigenvalue problem~(GEVP)
\begin{equation}\label{gevp}
\mathbb{C}_{\rm 2pt}(t) V(t, t_0)
=
\mathbb{C}_{\rm 2pt}(t_0) V(t, t_0) \Lambda(t, t_0) 
\end{equation}
for different reference times $t_0$ and $t>t_0$. This gives the matrix
of generalised eigenvalues $\Lambda(t, t_0) = \mathrm{diag}\left(
\lambda^\alpha(t, t_0) \right)$ and eigenvectors $V(t, t_0) =
(v^\alpha_i(t, t_0))$, where the superscripts~(subscripts) refer to
the eigenstate~(operator) and we employ the normalisation
$v^{\alpha\intercal} \mathbb{C}_{\rm 2pt}(t_0)v^\beta =
\delta^{\alpha\beta}$.  In the limit of large times, the eigenvalues
decay exponentially with the energy of the state,
$\lambda^\alpha(t,t_0) \propto e^{-E_\alpha (t-t_{0})}$, while the
elements of the eigenvectors $v^\alpha_i(t, t_0)$ are related to the
overlap of the operator $\oo_i $ with the state $\alpha$.

The effective energies $E_{\rm 
 eff}^{\alpha}(t)=a^{-1}\ln[\lambda^{\alpha}(t)/\lambda^{\alpha}(t+a)]$
are shown in Fig.~\ref{gevp_results} for $t_0=0.3~\mathrm{fm}$.\footnote{This 
value of $t_0$ proved the best choice for our setup in terms of
resolving the lowest two states and minimizing the statistical noise.}
For both bases, the lowest energy coincides with the nucleon mass on this
ensemble~\cite{RQCD:2022xux}, while the second level is close to the
sum of the nucleon and sigma energies. Therefore, we will identify
$\alpha=1$ with $N$ and $\alpha = 2$ with $N\sigma$. The third
eigenvalue for basis $\mathbb{B}_3$ does not exhibit a clear plateau,
with the statistical errors increasing rapidly for $t-t_0-a>3a$. We
were not able to obtain a reliable estimate of the $\sigma_8$ meson
mass, however, assuming the masses of the $\sigma_8$ and the physical
$f_0(980)$ are similar~(as is the case for $\sigma_0$ and $f_0(500)$),
the $N\sigma_8$ non-interacting level would lie around $2200$~MeV, 
above the almost degenerate lowest P-wave $N\pi$ and S-wave $N\pi\pi$ levels.

We observe that $E_{\rm eff}^1(t)$ is consistent with the effective
mass of the standard nucleon two-point function, $C_{\rm 2pt}^{\rm
  STD}(t)=\mathbb{C}_{\rm 2pt}(t)_{11}=\langle
\oo_p(t)\oob_p(0)\rangle$, see Fig.~\ref{gevp_results}, which plateaus
already at $t=0.5$~fm~(equal to $t-t_0-a=a$ for $t_0=3a$). 

The multi-particle interpolators contribute two orders of magnitude
less to $v^1$ than $\oo_p$, however, the impact on three-point
functions can still be significant, see below. Note that the $\oo_p$
contribution also dominates $v^2$ (and $v^3$).

\begin{table*}[t]
	\caption{\label{tab:fits} Results for the unrenormalized
          nucleon scalar charges from the standard ratio $R^{\rm
            STD}_{u+d,s}(t,\tau)$. These and the summed
          ratios~\eqref{summation} are fitted according to
          Eq.~\eqref{global_fit1} and
          Eqs.~\eqref{fit2_sum}~--~\eqref{fit1_sum}, respectively.
          For the latter, the fit range is varied for each model and a
          weighted average is quoted, along with the range of the
          $\chi^2/\mathrm{N}_{\mathrm{df}}$. Where applicable, the energy gaps
          $\Delta E$ between the first excited state and the ground
          state are given. Priors for $\Delta E$ based on the
          non-interacting S-wave $N\sigma$ or P-wave $N\pi$ energies
          are labelled as $\Delta E_{N\sigma}$ and $\Delta E_{N\pi}$,
          respectively, see the text.}  \centering
        \begin{ruledtabular}
	\begin{tabular}{p{4.5cm}>{\centering\arraybackslash}p{2.cm} >{\centering\arraybackslash}p{2.cm} >{\centering\arraybackslash}p{2.cm}> {\centering\arraybackslash}p{2.cm} >{\centering\arraybackslash}p{2.cm} >{\centering\arraybackslash}p{2.cm}} 
                		\textbf{Fit models} & \boldmath{$g^{{\rm lat},u+d}_S$} & \boldmath{$\Delta E$} \textbf{[MeV]} & \boldmath{$\chi^2\bigr/\mathrm{N}_{\mathrm{df}}$} & \boldmath{$g^{{\rm lat},s}_S$} & \boldmath{$\Delta E$} \textbf{[MeV]} & \boldmath{$\chi^2\bigr/\mathrm{N}_{\mathrm{df}}$}\\
                \midrule
                \textbf{Eq.~\eqref{fit2_sum}} | $c_{11}=0$ & $31.7 \pm 2.6$ & $495\pm 191$ & $[1.1,1.3]$ & $11.8 \pm 1.0$ & $526 \pm 193$ & $[1.1,1.3]$ \\
                \textbf{Eq.~\eqref{fit2_sum}} | $c_{11}=0$, $\Delta E_{N\sigma}$ & $30.8\pm 1.8$ & $516\pm 35$ & $[1.1,1.4]$  & $11.7 \pm 0.9$ & $514 \pm 33$ & $[1.1,1.3]$\\
                \textbf{Eq.~\eqref{fit2_sum}} | $c_{11}=0$, $\Delta E_{N\pi}$  & $29.0 \pm 1.6$ & $678 \pm 31$ & $[1.2,1.5]$  & $10.9 \pm 0.8$ & $678 \pm 30$ & $[1.2,1.4]$\\
        \textbf{Eq.~\eqref{fit1_sum}}  & $30.0 \pm 2.0$ & $-$ & $[1.0, 1.4]$ & $11.6 \pm 0.9$ & $-$ & $[0.9,1.5]$\\
                \midrule
                        \textbf{Eq.~\eqref{global_fit1}} & $31.8 \pm 2.6$ & $566 \pm 132$ & $0.63$& $12.3 \pm 0.9$ & $537 \pm 114$ & $0.61$\\
                        \textbf{Eq.~\eqref{global_fit1}} | $\Delta E_{N\sigma}$ & $31.6 \pm 2.0$ & $553 \pm 52$ & $0.63$& $12.2 \pm 0.7$ & $527 \pm 40$ & $0.61$\\
                        \textbf{Eq.~\eqref{global_fit1}} | $\Delta E_{N\pi}$ & $29.5 \pm 1.6$ & $690 \pm 39$ & $0.65$& $11.2 \pm 0.8$ & $686 \pm 32$ & $0.64$\\
        \end{tabular}
        \end{ruledtabular}
\end{table*}

An operator with an improved overlap to the ground state can be
constructed from a linear combination of the initial operators $\oo_i$
and the eigenvectors for the first level
$v^{\alpha=1}_i(t^\prime,t_0)$ evaluated at a large enough time
$t^\prime$ such that the eigenvector is stable~($t\ge t^\prime=6a$ for $t_0=3a$ in our setup),
\begin{equation}
\label{gevp_op}
\oo^{{\rm imp}}(t)\propto \sum_{i\in \mathbb{B}} v_i^{\alpha=1}({t}^\prime, t_{0}) \oo_i(t-t_0/2),
\end{equation}
with $\mathbb{B}=\mathbb{B}_2$ or $\mathbb{B}_3$.
This operator can be used to define GEVP-improved two- and three-point functions
\begin{align}\label{gevp_c3pt}
C^{{\rm imp}}_{\rm 2pt}(t) &= \langle \oo^{{\rm imp}}(t) ~\oob^{{\rm imp}}(0) \rangle,\\
C^{{\rm imp},\mathcal{S}^q}_{\rm 3pt}(t, \tau) &= \langle \oo^{{\rm imp}}(t) ~\mathcal{S}^q(\tau) ~\oob^{{\rm imp}}(0) \rangle.
\end{align}

We neglect the $N\sigma \stackrel{\mathcal{S}^q}{\longrightarrow} N\sigma$ contributions to  $C_{\rm 3pt}^{{\rm imp}, \mathcal{S}^q}$ 
that are suppressed by the second power of the small eigenvector component 
$v^1_{p\sigma}$ for $\mathbb{B}_2$~($v^1_{p\sigma_0, p\sigma_8}$ for $\mathbb{B}_3$). 
In particular, we do not expect the matrix element $\langle N\sigma|\mathcal{S}^q|N\sigma \rangle$ 
to be enhanced relative to $\langle N|\mathcal{S}^q|N\rangle$.
In order to assess the size of excited state contamination to the
improved three-point function we form the ratio
\begin{equation}\label{gevp_ratios}
R^{\mathrm{GEVP}}_{u+d,s}(t, \tau) = \frac{C^{{\rm imp},\mathcal{S}^{u+d,s}}_{\rm 3pt}(t,\tau)}{C^{{\rm imp}}_{\rm 2pt}(t)} 
\end{equation}
for the $\mathcal{S}^{u+d}$ and $\mathcal{S}^{s}$ currents for 10
source-sink separations, $t=2a-11a=0.2-1.1$~fm, shown in
Fig.~\ref{fig:gevp_std_ratios}. Considering the spectral decomposition
given in Sec.~\ref{sec:method}, this ratio tends to the charges
$g_S^{{\rm lat},u+d}$ and $g_S^{{\rm lat},s}$ of the nucleon, respectively, at
large time separations. For comparison we also evaluate
$R^{\mathrm{STD}}_{u+d,s}(t, \tau)$, the ratio of the standard
three-point function $C_{\rm 3pt}^{\mathrm{STD}}(t,\tau)=\langle \oo_p(t)
\mathcal{S}^{u+d,s}(\tau)\oob_p(0)\rangle$ to
$C_{\rm 2pt}^{\mathrm{STD}}(t)$ for $t=2a-17a=0.2-1.7$~fm. Note that only
the ratios up to $t=14a$ are shown in the figure.  The standard ratios
show a significant dependence on the source-sink separation and
current insertion time, indicating large excited state contributions
to the three-point function. In contrast, the GEVP ratios display a
much milder dependence. This suggests that the S-wave $N\sigma$
contributions to $C_{\rm 3pt}^{\mathrm{STD}}(t,\tau)$ are significant and 
that these are effectively removed when employing the GEVP-improved
interpolator.  However, clearly, residual excited state contamination
remains and a larger basis of operators would need to be considered to
reduce these contributions further.

\section{Fitting analysis}\label{sec:fit}

In the following, we present the fitting analysis carried out to
extract the scalar charges from the ratios of the three-point to two-point
correlation functions. With a large
number of source-sink separations~($t$) at our disposal, we employ the
summation method and compute
\begin{equation}\label{summation}
R^{\rm sum}_{u+d,s}(t) = \sum_{\tau= +c}^{t-c} R_{u+d,s}(t,\tau),
\end{equation}
where the sum over the current insertion~($\tau$) runs from time slice $\tau=c$
up to $t-c$.  Considering the spectral decompositions in
Eqs.~\eqref{eq:specdecomp2} and~\eqref{eq:specdecomp3} truncated after
the first excited state, the $t$-dependence of the summed ratio reads
\begin{align}
&  R^{\rm sum}_{u+d,s}(t) = g^{{\rm lat},u+d,s}_S(t-2c+a) +\nonumber\\
 &  2c_{10}~\frac{e^{\Delta E(c-t)} - e^{\Delta E(a-c)}}{1-e^{\Delta E}}
+c_{11}\left(t-2c+a \right)e^{-\Delta E t}.\label{fit2_sum}
\end{align}
The coefficients $c_{10}$ and $c_{11}$ are related to the matrix
elements $\langle N|\mathcal{S}^{u+d,s}|2\rangle$ and $\langle
2|\mathcal{S}^{u+d,s}|2\rangle$, respectively, and $\Delta E$ denotes
the energy gap between the ground $|N\rangle = |1\rangle$ and the
first resolvable excited state $|2\rangle$.  In the limit of ground state
dominance, this dependence simplifies to
\begin{equation}
R^{\rm sum}_{u+d,s}(t) = c_0 + g^{{\rm lat},u+d,s}_St\label{fit1_sum}
\end{equation}
with $c_0=g_S^{u+d,s}(a-2c)$.

First, we discuss the fits to the standard ratios $R^{\rm
  STD}_{u+d,s}(t,\tau)$. Given the significant excited state
contributions for smaller $t$ seen in Fig.~\ref{fig:gevp_std_ratios},
we perform correlated fits to the summed ratios employing
Eq.~\eqref{fit2_sum} with $c=2a$. The latter means that the summed
ratio is realised for $t\ge 4a$. We set $c_{11}=0$ as the term
involving the excited-state to excited state transition was not
resolved. The fit range is varied and we take a weighted average of
the results.\footnote{The weighted average $\bar{a}$ of fit
coefficient $a\in\{g_S^{u+d,s},\Delta E\}$ is computed using weights
$w_i=N\mathrm{exp}\left(-(\chi_i^2 - \mathrm{N}_{\mathrm{df}i})/2\right)$, where
$\chi^2_i$ and $\mathrm{N}_{\mathrm{df}i}$ are the fit quality and number of
degrees of freedom for fit $i$, respectively. $N$ is chosen so that
$\sum_iw_i=1$. The average $\bar{a}$ and its uncertainty $\Delta a$
are given by $\bar{a}=\sum_i w_i a_i$ and $\Delta a^2=\sum_i w_i
\Delta a_i^2$, where $\Delta a_i$ is the bootstrapped error for fit
$i$.} The values obtained for $g_S^{u+d,s}$ and $\Delta E$ are listed
in Tab.~\ref{tab:fits}. The energy gap is not well determined in the
fits, however, we know from the GEVP analysis~(see
Fig.~\ref{gevp_results}) that the first excited state energy is
consistent with the S-wave $N\sigma$ non-interacting level. This
motivates us to repeat the analysis imposing a prior on $\Delta E$
equal to the mass of the sigma meson with a width of 50~MeV~($9\%$),
leading to similar results for the scalar charges but with smaller
uncertainties.  In the absence of detailed knowledge of the energy
spectrum, a prior for the energy gap is often set using the P-wave
$N\pi$ energy~(although additional excited states are usually taken
into account in the fit model). A larger energy gap gives consistent
but slightly lower values for the scalar charges. For $t\geq 8a$ we
are also able to perform linear fits~\eqref{fit1_sum} (see
Tab.~\ref{tab:fits}) since the excited state contributions to $R^{\rm
  sum}_{u+d,s}$ fall below the noise.

For completeness, we also carried out correlated fits to
$R^{\mathrm{STD}}_{u+d,s}(t, \tau)$ with the following ansatz:
\begin{equation}\label{global_fit1}
R_{u+d,s}(t,\tau)  = g^{{\rm lat},u+d,s}_S + c_{10}\left(e^{-\Delta E (t-\tau)} + e^{-\Delta E \tau}\right),
\end{equation}
where, again, the excited-state to excited-state term is omitted. We
simultaneously fit the ratios with $t/a=8,9,11,13,15$ employing the fit
range $2a\le \tau \le t-2a$ for each $t$.\footnote{For our setup,
with 800 configurations employed in the analysis, this choice ensures
that the covariance matrix is well determined.} The largest source-sink
separations are not included in the fit due to the deterioration in
the signal~(note that the summed ratios display smooth behaviour up to
$t=17a$). The results, detailed in Tab.~\ref{tab:fits}, are compatible
with those for the summed ratios.

Turning to the GEVP-improved summed ratios~(again with $c=2a$, $t\ge 4a$), we
find that the excited state contributions are sufficiently suppressed
to employ the linear fit ansatz~\eqref{fit1_sum} starting from
$t=4a$. The final results for the weighted averages for basis
$\mathbb{B}_2$ are
\begin{align}
  g^{{\rm lat},u+d}_S&=30.1\pm1.8 \qquad [1.0,0.8,1.0] \nonumber\\
  g^{{\rm lat},s}_S&=11.7\pm0.7 \qquad [1.0,1.2, 1.3]. \label{gS_GEVP_B2}
\end{align}
The $\chi^2/\mathrm{N}_{\mathrm{df}}$ for the three fits which enter the
average~(with fit ranges $t/a=4-11$, $t/a=5-11$ and $t/a=6-11$)
are given in brackets. For $\mathbb{B}_3$, we obtain
\begin{align}
  g^{{\rm lat},u+d}_S&=30.3\pm1.8  \qquad [1.1, 0.8, 1.1] \nonumber\\
  g^{{\rm lat},s}_S&=11.6\pm0.9  \qquad [1.1, 0.9, 1.2] \label{gS_GEVP_B3}
\end{align}
for the same fit ranges. These four fit results for the charges are
displayed as the green bands in Fig.~\ref{fig:gevp_std_ratios}.  The
standard and GEVP summed ratios are shown in
Fig.~S3 of the Supplemental Material, along with
representative fits utilising Eqs.~\eqref{fit2_sum} and \eqref{fit1_sum}.

\section{Summary and discussion}\label{sec:conclude}
We utilised the variational approach with a basis including
nucleon-sigma type interpolators to remove the dominant excited state
contamination from the three-point correlation functions with a scalar
current. The scalar charges are reliably extracted from source-sink
separations $t\le1.1$~fm. The results of the standard
method~(Tab.~\ref{tab:fits}) agree well with the GEVP-improved
numbers~\eqref{gS_GEVP_B2} and~\eqref{gS_GEVP_B3}, however, the former
require larger time separations~(see also the comparison in
Fig.~\ref{fig:gevp_std_ratios}). Considering the exponential increase
in the noise over signal with $t$, we find the new method to be cost
effective.  Converting our values via Eqs.~\eqref{eq:sigm1}
and~\eqref{eq:sigm2} into the $\sigma$-terms, we obtain $\sigma_{\pi
 N}=(235\pm 24)$~MeV and $\sigma_{sN}=(42\pm 16)$~MeV, where the
large uncertainty, in particular for $\sigma_{sN}$, is mostly due to
the large value of $r_m$ at our coarse lattice spacing $a\approx
0.098$~fm ($r_m\approx3.4$~\cite{Heitger:2021bmg}). The results are
consistent with those at similar pion masses from other
works~\cite{Yang:2015uis,Yamanaka:2018uud,Petrak:2023qhx}.

For physical quark masses, the $\sigma$ meson becomes unstable and one
might be tempted to perform a full scattering
analysis~\cite{PhysRevLett.118.022002,Liu:2016cba,Fu:2017apw,Guo:2018zss,RBC:2023xqv,Rodas:2023gma,Bruno:2023pde},
including $N\pi\pi$ type interpolators. However, since the width of
the $f_0(500)$ is similar to its
energy~\cite{ParticleDataGroup:2024cfk}, also a scalar bilinear
interpolator will most likely couple well to the scattering
states~\cite{Lang:2016hnn,Kiratidis:2016hda,Severt:2022jtg}, as well
as to the local isoscalar current. Therefore, the method presented
here may very well be directly applicable to the physical
point. Future studies with closer to physical pion masses will shed
further light on this.

\begin{acknowledgments}
We thank P.~Petrak for sharing results on the sigma terms with us.
L.~B.\ is grateful to his colleagues at DESY and at the Humboldt
University for stimulating discussions, with special thanks to
J.\ Green and R.\ Sommer for their insightful comments.  He also
extends his gratitude to R.\ Gupta and K.-F.\ Liu and the members of
the $\chi$QCD collaboration for useful discussions.  G.~B.\ thanks
S.\ \mbox{Leupold} for discussions. L.~B. received support through the German
Research Foundation (DFG) research unit FOR5269 ``Future methods for
studying confined gluons in QCD''.  Simulations were performed on the
QPACE~3 computer of SFB/TRR\nobreakdash-55, using an adapted version of
the {\sc Chroma}~\cite{Edwards:2004sx} software package.
\end{acknowledgments}
\bibliography{bibliography}
\newpage
\appendixsection
\section*{Construction of the two- and three-point functions}\vspace{-0pt}

\begin{figure}[ht]
	\centering
	\includegraphics[width=\columnwidth]{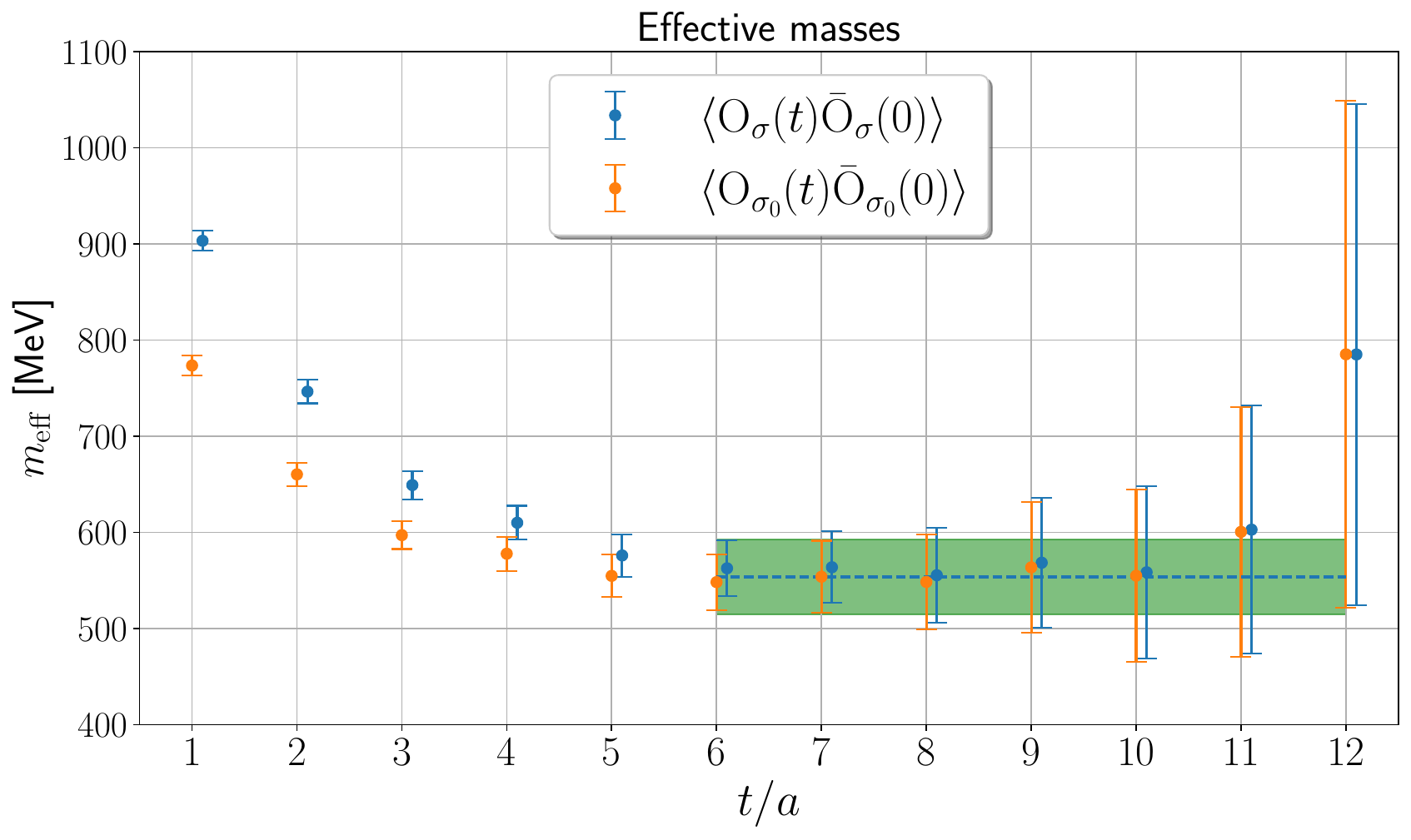}
	\caption{Effective masses of the two-point correlation
          functions constructed from the interpolators
          $\oo_{\sigma}$~(orange points) and $\oo_{\sigma_0}$~(blue
          points). The blue dashed line and green band indicate the
          mass, $m_\sigma =\left(554 \pm 39\right) {\rm MeV}$,
          extracted from a fit to $\langle
          \oo_{\sigma_0}(t)\oob_{\sigma_0}(0)\rangle$ in the range
          $t/a=6-12$.}
	\label{fig:effective_mass_sigma}
\end{figure}

\begin{figure*}[ht]
	\centering
	\begin{subfigure}{0.23\textwidth}
		\includegraphics[width=\textwidth]{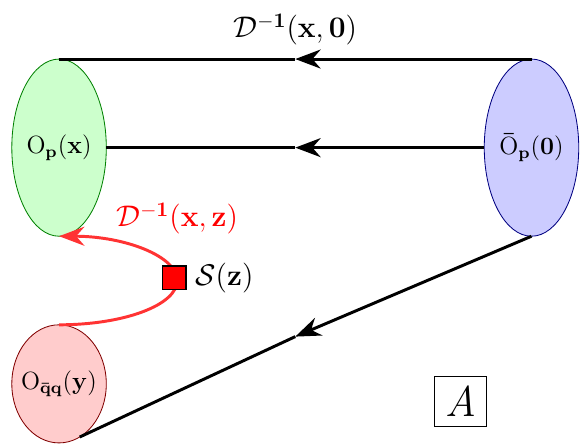}
	\end{subfigure}
	\hspace{0.8em}
	\begin{subfigure}{0.23\textwidth}
		\includegraphics[width=\textwidth]{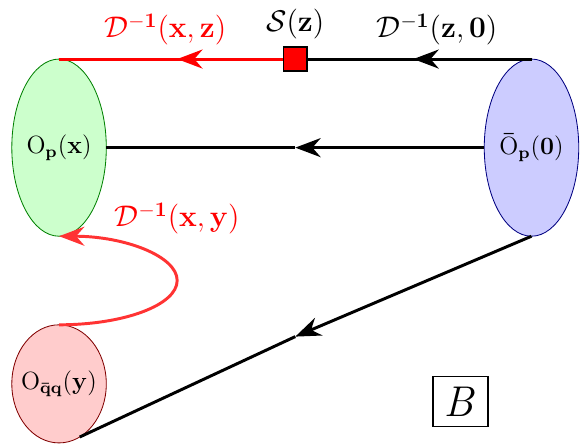}
	\end{subfigure}
	\hspace{0.8em}
	\begin{subfigure}{0.23\textwidth}
		\includegraphics[width=\textwidth]{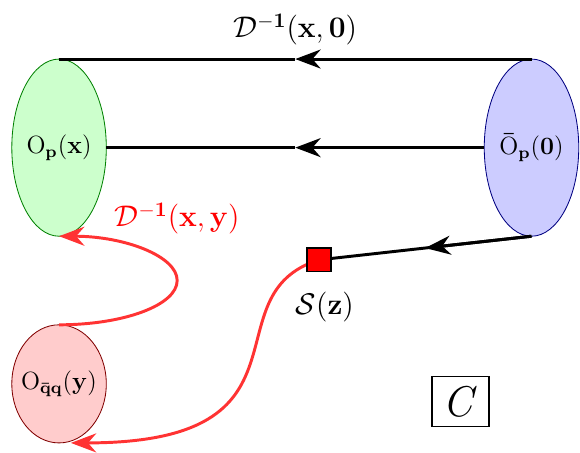}
	\end{subfigure}
	\hspace{0.8em}
	\begin{subfigure}{0.23\textwidth}
		\includegraphics[width=\textwidth]{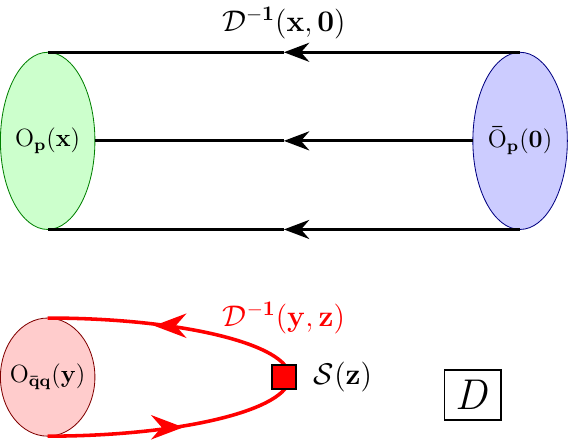}
	\end{subfigure}
	\hfill
	\vspace*{2em}	
	\begin{subfigure}{0.23\textwidth}
		\includegraphics[width=\textwidth]{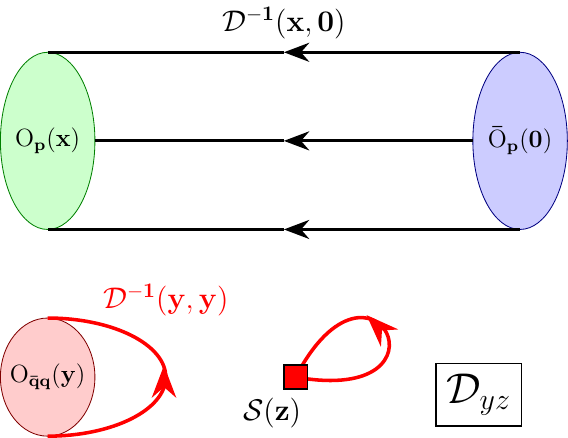}
	\end{subfigure}
	\hspace{2em}
	\begin{subfigure}{0.23\textwidth}
		\includegraphics[width=\textwidth]{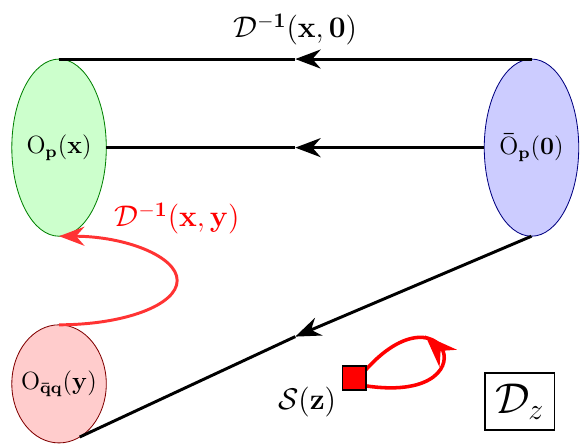}
	\end{subfigure}
	\hspace{2em}
	\begin{subfigure}{0.23\textwidth}
		\includegraphics[width=\textwidth]{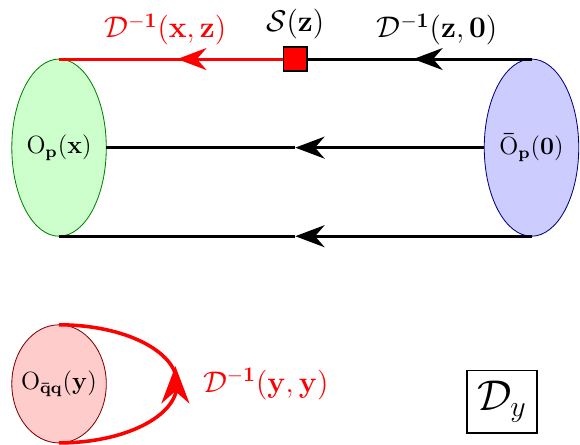}
	\end{subfigure}
\caption{Topologies in position space arising from
	the Wick contractions of the three-point correlation functions
	$\langle \mathrm{O}_{N\sigma}(x, y) ~\mathcal{J}(z)
	~\bar{\mathrm{O}}_{N}(0) \rangle$. The three black lines
	represent point-to-all propagators, the two red lines in each topology 
	correspond to all-to-all propagators, and the red square indicates the current
	insertion at $z=(\mathbf{z},\tau)$. The $\mathrm{O}_{N\sigma}(x, y)$
	interpolator contains a $qqq$-structure $\mathrm{O}_{p}(x)$ at
	the spacetime position $x=(\mathbf{x}, t)$ and a
	$\bar{q}q$-structure $\mathrm{O}_{\bar{q}q}(y)$ at
	$y=(\mathbf{y}, t)$.}
\label{fig:diagrams_nJnsigma}
\end{figure*}

We aim to improve the extraction of the scalar matrix element of the
nucleon at rest by constructing improved three-point correlation
functions~\eqref{gevp_c3pt} via the variational approach utilising
bases including the standard proton interpolator~(with a $qqq$
structure) and S-wave proton-sigma interpolators~(with a
$qqq-\bar{q}q$ structure). For the latter, we combine a standard
proton and a (scalar) sigma operator that are both projected onto zero
momentum. For the two bases realised in the study~(see
Sec.~\ref{sec:variat}), we implement the proton
operator~($\oo_p(\zzero)$) and three multi-particle interpolators
\begin{align}
\oo_{p\sigma} &= \oo_p(\zzero) \oo_\sigma(\zzero),\\
\oo_{p\sigma_0} &= \oo_p(\zzero) \oo_{\sigma_0}(\zzero),\\
\oo_{p\sigma_8} &= \oo_p(\zzero) \oo_{\sigma_8}(\zzero).
\end{align}
The flavour content of the sigma operators corresponds to the
SU(2) singlet~($\oo_{\sigma}$) and SU(3) singlet and
octet~($\oo_{p\sigma_0}$ and $\oo_{p\sigma_8}$, respectively),
combinations.

To improve the overlap with the lowest lying levels, the proton and
sigma interpolators in the single and multi-particle operators are
spatially extended. For the former, we apply the same quark field
smearing as employed in \cite{Barca:2024sub}, i.e., 150 Wuppertal
smearing steps\footnote{The gauge links are APE
smeared~\cite{Falcioni:1984ei}, see~\cite{RQCD:2022xux} for details.} with $\alpha=0.25$,
corresponding to a smearing radius of $\langle r^2\rangle^{1/2}\sim
1.0$~fm for the proton interpolator.  Regarding the sigma meson
interpolators, we investigated 5 different smearings:
$0, 5, 20, 40$ and $150$ iterations.
We find that $5$ iterations, with a $\langle r^2\rangle^{1/2}\sim
0.2$~fm for the interpolator, is sufficient to reduce excited state
contributions to the two-point functions $C_{\rm
  2pt}^{\sigma\sigma}(t)=\langle \oo_\sigma(t)\oob_\sigma(0)\rangle$
and $C_{\rm 2pt}^{\sigma_0\sigma_0}(t)=\langle
\oo_{\sigma_0}(t)\oob_{\sigma_0}(0)\rangle$ such that the respective
effective masses plateau around $t=6a=0.6$~fm, see
Fig.~\ref{fig:effective_mass_sigma}. Fitting to $C_{\rm
  2pt}^{\sigma_0\sigma_0}(t)$ in the range $t/a=6-12$ we obtain
$m_\sigma =\left(554 \pm 39\right) {\rm MeV}$.\footnote{We remark that 
simulations away from the flavour symmetric 
point~\cite{PhysRevLett.118.022002,Briceno:2017qmb} at $M_\pi=391$~MeV 
and $M_K=549$ MeV find a much larger mass $m_\sigma=745(5)$~MeV.} 
A higher number of iterations increases the noise without further reducing the
excited state contamination. Note that, as we are working in
the SU(3) symmetric limit, there is no mixing between the singlet and
octet flavour sectors. However, the spectrum of states created by the SU(2)
singlet operator $O_\sigma$ will include both singlet and octet
levels. Concerning the $\sigma_8$ interpolator, we were not able to
achieve a clear plateau in the respective two-point function for any
of the 5 smearings realised.

\begin{figure*}[ht]
  \centering
  \begin{subfigure}{0.48\textwidth}
    \includegraphics[width=\columnwidth]{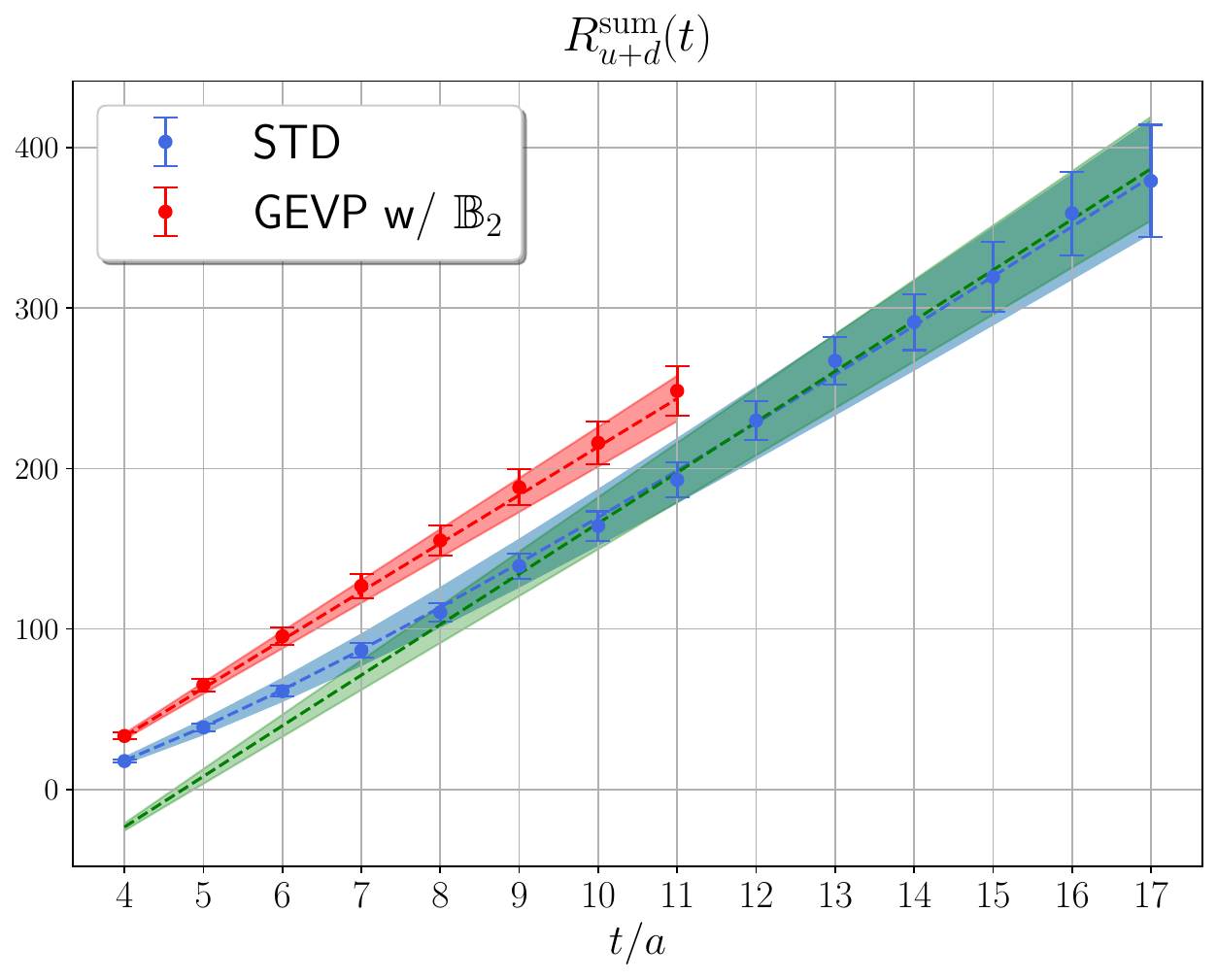}
  \end{subfigure}
  \hspace{0.8em}
  \begin{subfigure}{0.48\textwidth}
    \includegraphics[width=\columnwidth]{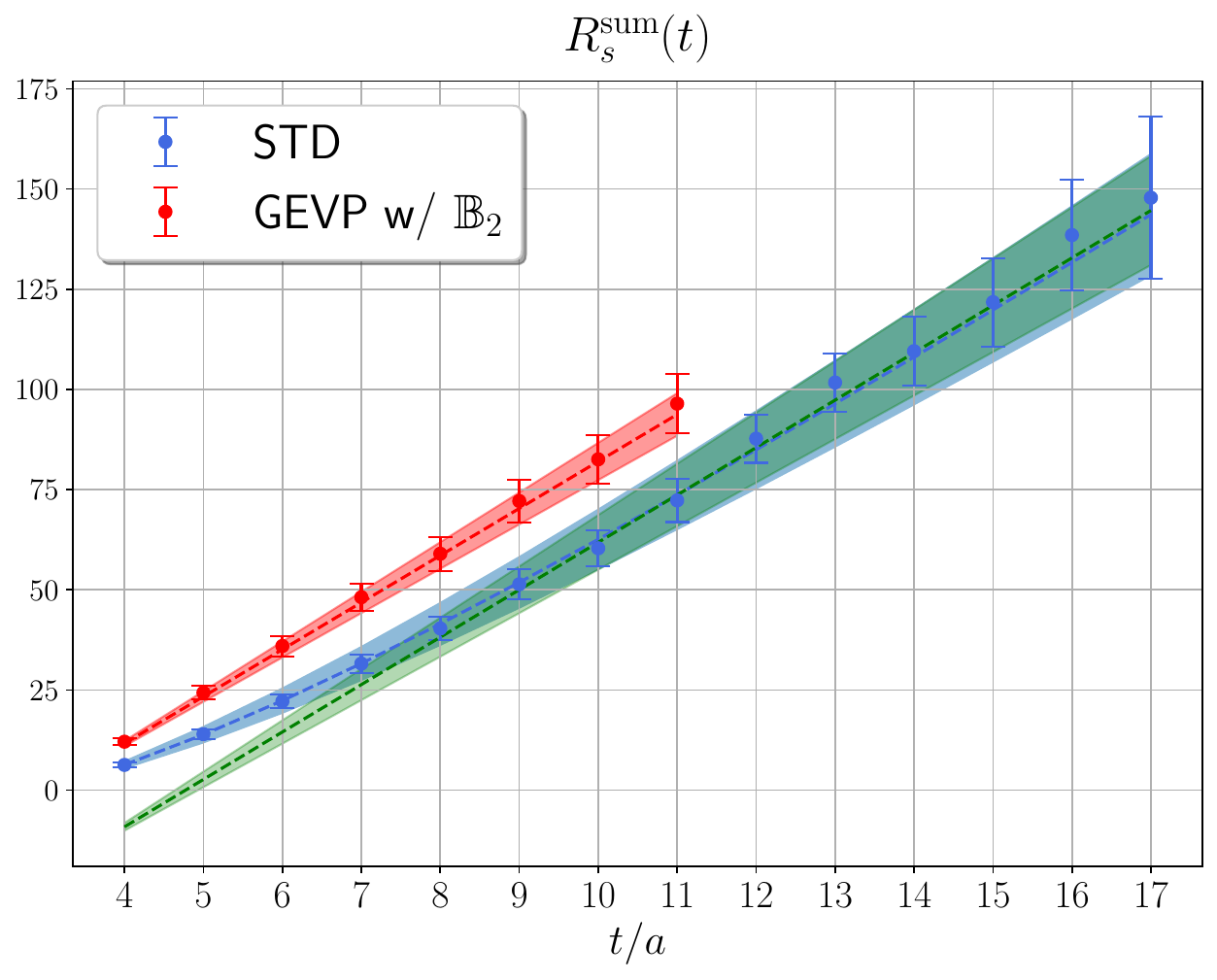}
  \end{subfigure}
  \caption{\label{fig:summed_ratios}
    Results for the summed ratios, Eq.~\eqref{summation}
    with $c=2a$, for the $u+d$ (left) and $s$ (right) flavour
    combinations of the scalar current.  The GEVP-improved
    results~(red points) are obtained utilising basis
    $\mathbb{B}_2$. The red dashed line and band indicate the
    central value and uncertainty of a linear
    fit~\eqref{fit1_sum} over the range $t/a=4-11$. For the
    standard summed ratios~(blue points) a linear fit for
    $t/a=10-17$~(light green band) is displayed along with a fit
    including contributions from the first excited state~(blue
    band) (\eqref{fit2_sum} with $c_{11}=0$) for $t/a=4-17$.}
\end{figure*}

In order to implement the variational approach, the two-point functions
in Eq.~\eqref{gevp_c2pt} and the following three-point functions
\begin{align}
	\label{std_3pt}
	&\langle
	\mathrm{O}_{p}(t)
	~\mathcal{S}^q(\tau)
	~\bar{\mathrm{O}}_{p}(0)
	\rangle~,      \\
	\label{p2psigma_3pt}
	&\langle
	\mathrm{O}_{p\sigma}(t)
	~\mathcal{S}^q(\tau)
	~\bar{\mathrm{O}}_{p}(0)
	\rangle~, \\
	\label{p2psigma0_3pt}
	&\langle
	\mathrm{O}_{p\sigma_0}(t)
	~\mathcal{S}^q(\tau)
	~\bar{\mathrm{O}}_{p}(0)
	\rangle~, \\
	\label{p2psigma8_3pt}
	&\langle
	\mathrm{O}_{p\sigma_8}(t)
	~\mathcal{S}^q(\tau)
	~\bar{\mathrm{O}}_{p}(0)
	\rangle~,
\end{align}
need to be evaluated in the rest frame of the proton. We remind the
reader that we neglect the $N\sigma
\stackrel{\mathcal{S}^q}{\longrightarrow} N\sigma$ contributions to
the GEVP-improved three-point functions~\eqref{gevp_c3pt}. In the
following, we restrict the discussion to the calculation of the
correlation functions involving the proton-sigma
interpolators. Furthermore, as the relevant two-point functions can be
constructed from \eqref{std_3pt}-\eqref{p2psigma8_3pt} replacing the
scalar current with smeared scalar operators inserted at $\tau=0$, we
only consider the three-point functions
\eqref{p2psigma_3pt}-\eqref{p2psigma8_3pt}.
Figure~\ref{fig:diagrams_nJnsigma} displays the quark-line diagram
topologies which arise from the Wick contractions of these three-point
functions.  The connected diagrams A, B, and C and the connected parts
involving the proton interpolator of the remaining diagrams are
constructed from point-to-all and sequential propagators.  Stochastic
estimation is employed to compute the quark loops for diagrams
$\mathcal{D}_{yz}$, $\mathcal{D}_{z}$, and $\mathcal{D}_y$ and the
one-end trick is utilised to determine the local to smeared scalar
two-point function in diagram D. To improve the determination of this
diagram and also diagrams $\mathcal{D}_{yz}$ and $\mathcal{D}_y$, we
realise four source positions for the scalar two-point function and
the proton two- and three-point functions. In total, we require the
equivalent of $\approx 22$ propagators for each source-sink
separation. However, in order to investigate the efficacy of the
variational approach and, in particular, to ensure a reliable
extraction of the corresponding eigenvectors, we realise source-sink
separations up to $t=11a=1.1$~fm. When taking the recycling of
propagators into account, this means that we need
$\approx 132$ propagators overall per configuration for the GEVP analysis.

Finally, we remark that as the scalar operator has the quantum numbers
of the vacuum, we implement the vacuum-subtracted current
$\widetilde{\mathcal{S}}^q=\mathcal{S}^q - \langle \mathcal{S}^q
\rangle$ and interpolators
$\widetilde{\oo}_\sigma={\oo}_\sigma- \langle {\oo}_\sigma \rangle$,
$\widetilde{\oo}_{\sigma_0}={\oo}_{\sigma_0}- \langle {\oo}_{\sigma_0} \rangle$ and
$\widetilde{\oo}_{\sigma_8}={\oo}_{\sigma_8}- \langle {\oo}_{\sigma_8} \rangle$.
Also within the multi-particle interpolators, we replace $\oo_{p\sigma}$ with
$\widetilde{\oo}_{p\sigma}={\oo}_{p}({\oo}_\sigma- \langle {\oo}_{\sigma} \rangle)$ etc.

\vspace{1em}
\section*{Summation method results}
In this section, we present results for the summed ratios $R^{\rm
  sum}_{u+d,s}$~\eqref{summation} for both the GEVP-improved and
standard correlation functions.  Figure~\ref{fig:summed_ratios}
displays the summed ratios obtained with $c=2a$, where basis
$\mathbb{B}_2$ is employed for the variational approach. Note that for
this value of $c$, $R^{\rm sum}_{u+d,s}$ is evaluated for $t\ge 4a$.
The results for basis $\mathbb{B}_3$ are consistent within errors.  As
seen in the figure, the standard $R^{\rm sum}_{u+d,s}$ suffers from
excited state contamination for $t < 8a=0.8$~fm. In contrast, for the
GEVP-improved $R^{\rm sum}_{u+d,s}$, the $N\sigma$ contributions at
short time separations are removed and ground state dominance sets in
already at $t\approx 4a=0.4$~fm. The fits performed reflect this
behaviour. For the standard summed ratios, contributions from the
ground state and a single excited state are modelled using
Eq.~\eqref{fit2_sum} when fitting in the range $t_{\rm min}$ to
$t_{\rm max}$ with $t_{\rm min}= 4a-6a$ and $t_{\rm max}=17a$. For
$t_{\rm min}\ge 8a$ a linear fit~\eqref{fit1_sum} is
sufficient. Similarly, a linear fit to the GEVP-improved $R^{\rm
  sum}_{u+d,s}$ for $t_{\rm min}\ge 4a$ with $t_{\rm max}=11a$ gives a
good fit quality. Representative examples of these fits are presented
in Fig.~\ref{fig:summed_ratios}. We remark that compatible values for
the scalar charges are obtained when fitting to the summed ratios with
$c=3a$.

\end{document}